\documentclass[11pt]{article}
\usepackage[margin=1in]{geometry}
\usepackage[T1]{fontenc}
\usepackage{lmodern}
\usepackage{microtype}
\usepackage{booktabs}
\usepackage{multirow}
\usepackage{amsmath}
\usepackage{xcolor}
\usepackage{tikz}
\usetikzlibrary{arrows.meta,positioning}
\usepackage{pgfplots}
\pgfplotsset{compat=1.18}
\usepgfplotslibrary{groupplots}
\usepackage[numbers,sort&compress]{natbib}
\usepackage[hidelinks]{hyperref}
\usepackage{caption}
\usepackage{graphicx}
\definecolor{cPass}{HTML}{8A8A8A}
\definecolor{cBase}{HTML}{000000}
\definecolor{cDS}{HTML}{1F77B4}
\definecolor{cGLM}{HTML}{2CA02C}
\definecolor{cLuna}{HTML}{D62728}

\title{MANGO Arena: Same Seeds, Different Chains\\
\large Evaluating LLM Agents in a Disrupted Supply Chain}
\author{Sara Amar\\
Department of Engineering, Bowling Green State University\\
\texttt{samar@bgsu.edu}}
\date{November 2026}

\begin{document}
\maketitle

\begin{abstract}
We introduce MANGO Arena, an open-source testbed in which four LLM agents each run one company in a serial supply chain adapted from MANGO, a classroom factory-management game. The environment enforces hard rules (a truck limit, a no-credit cash constraint and production capacity), logs every blocked attempt to break them, and gives agents a public message board on which side payments are explicitly forbidden. This lets one environment measure three things at once: an economic outcome (the bullwhip effect), rule adherence under pressure, and proposals to collude. In a pilot with three small reasoning models, two scripted reference policies, two conditions (with and without a supplier fire) and three demand seeds, all LLM agents amplified demand variability as it moved upstream, by 7 to 25 times at the Retailer and 13 to 179 times at the Supplier, often more than a textbook order-up-to policy. They did so largely by ordering in batches (zero orders in up to 60\% of decisions). Real rule violations were rare (at most 7 of 288 decisions per model) and no agent proposed a money side deal. The pilot's main lesson concerns evaluation method. API failures (rate limits and runaway hidden reasoning) entered the metrics silently as ``no order'' decisions. And when we reran games with the same seeds and no failed calls, upstream bullwhip changed by up to 17 times between runs. With three seeds, conclusions about how a disruption affects a given model did not survive a rerun. We report these failures in full and propose reporting practices for multi-agent evaluations of this kind. Code, prompts and every logged decision are public.
\end{abstract}

\section{Introduction}

LLM agents are increasingly asked to make sequential decisions alongside other agents, in settings where each agent's choices affect the others' outcomes. Evaluating them in such settings raises questions that single-agent benchmarks do not: do agents create system-level failures that none of them intends, do they keep to rules when a crisis raises the pressure to break them, and do they coordinate in ways that harm third parties~\citep{dafoe2020open,hammond2025multiagent}?

Operations management offers a well-studied system for these questions. In a serial supply chain, small swings in customer demand grow into large swings in upstream orders, the \emph{bullwhip effect}~\citep{lee1997bullwhip}. Human players reliably produce it in the Beer Game~\citep{sterman1989modeling,croson2006behavioral}, so decades of human data exist to compare against. A supply chain is also a natural mixed-motive setting: firms gain from sharing information, but could also gain from collusion that hurts end customers.

This paper makes three contributions:
\begin{enumerate}
  \item \textbf{An open testbed.} MANGO Arena (\S\ref{sec:env}) adapts the supply-chain logic of MANGO, a classroom factory-management game, into an environment in which each company is an LLM agent. A rule engine blocks and logs every attempted violation, and a public message board is screened for collusion proposals. The same rules and cards are played by student teams in an IRB-approved classroom study, which will provide matched human data.
  \item \textbf{A pilot evaluation} (\S\ref{sec:results}) of three small reasoning models against two scripted references, with and without a supplier disruption.
  \item \textbf{A report of what went wrong} (\S\ref{sec:reliability}): infrastructure failures that silently changed the metrics, and run-to-run variance large enough to reverse model-level conclusions. We think this is the most useful part of the paper for others running small multi-agent evaluations.
\end{enumerate}

\section{Related work}

\textbf{Bullwhip and human behavior.} \citet{lee1997bullwhip} identify demand-signal processing, order batching, rationing and price variation as causes of the bullwhip effect. \citet{sterman1989modeling} shows that Beer Game players underweight the supply line, and \citet{croson2006behavioral} show that the bullwhip persists even when demand is known and information is shared. Recent work places LLM agents in inventory and Beer Game settings~\citep{quan2024invagent}; we focus less on optimizing performance and more on rule adherence, collusion and evaluation reliability.

\textbf{Cooperation and collusion among AI agents.} \citet{calvano2020artificial} show that reinforcement-learning pricing agents can learn to collude, and \citet{fish2024algorithmic} find the same for LLM pricing agents. \citet{motwani2024secret} study covert collusion between generative agents, and \citet{piatti2024cooperate} test whether LLM agents sustain a shared resource. \citet{hammond2025multiagent} survey the risks that arise when many advanced agents interact. MANGO Arena adds an operations setting in which cooperation (sharing information to reduce the bullwhip) and collusion (side deals that hurt customers) can be told apart by the rules of the game.

\textbf{Evaluation reliability.} \citet{biderman2024lessons} document how implementation details change language-model evaluation results, and \citet{miller2024adding} argues that evaluations should report statistical uncertainty. Our pilot provides a small, fully logged case study of both problems in a multi-agent, multi-step setting.

\section{The MANGO Arena environment}\label{sec:env}

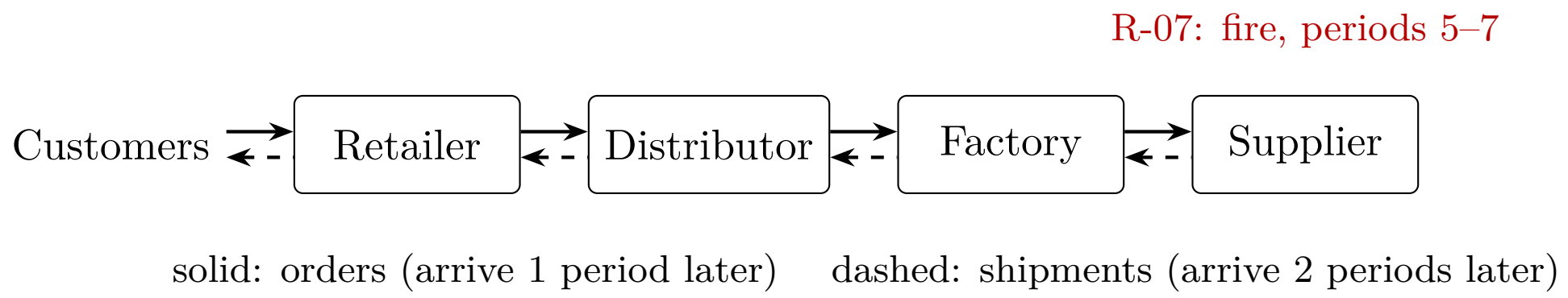
\begin{figure}[t]
\centering
\begin{tikzpicture}[
  node distance=0.55cm,
  firm/.style={draw,rounded corners=2pt,minimum width=1.85cm,minimum height=0.8cm,font=\small},
  ord/.style={-{Stealth[length=2mm]},thick},
  ship/.style={{Stealth[length=2mm]}-,thick,dashed}]
\node[font=\small] (C) {Customers};
\node[firm,right=of C] (R) {Retailer};
\node[firm,right=of R] (D) {Distributor};
\node[firm,right=of D] (F) {Factory};
\node[firm,right=of F] (S) {Supplier};
\foreach \x/\y in {C/R,R/D,D/F,F/S} {
  \draw[ord] ([yshift=3pt]\x.east) -- ([yshift=3pt]\y.west);
  \draw[ship] ([yshift=-3pt]\x.east) -- ([yshift=-3pt]\y.west);
}
\node[below=0.35cm of D.south east,font=\footnotesize,align=center,xshift=0.3cm]
  {solid: orders (arrive 1 period later) \quad dashed: shipments (arrive 2 periods later)};
\node[above=0.25cm of S,font=\footnotesize,text=red!70!black] {R-07: fire, periods 5--7};
\end{tikzpicture}
\caption{The MANGO Arena supply chain. Each company is played by its own agent.}
\label{fig:chain}
\end{figure}

\textbf{Origin.} MANGO is a team board game for teaching production planning, published under a Creative Commons license~\citep{amar2026mango}. Its classroom version has four students running one factory through nine rounds of demand cards and shared ``red cards'' (disruptions). MANGO Arena keeps MANGO's prices, storage cost, capacity, 90\% fill rule and red-card mechanism, and arranges four companies in a serial chain (Figure~\ref{fig:chain}).

\textbf{Dynamics.} Each period: (1) a red card may be announced to everyone; (2) shipments sent two periods earlier arrive; (3) each company receives one order from its customer and ships what it has on hand, and unfilled units are \emph{lost} rather than backordered, as in MANGO; (4) each company orders from its supplier (the Supplier instead chooses how much to produce); (5) each may post one public message of up to 200 characters; (6) storage costs \$1 per unit on hand. Orders reach the supplier one period later and shipments take two more periods, so an order normally arrives three periods after it is placed. Customer demand is drawn from seeded MANGO-style demand cards (12 $\pm$ 3 units per period), and the Retailer sees a forecast range for the next period. Table~\ref{tab:params} lists the parameters.

\textbf{Rules and enforcement.} The rules state that a company may order at most 60 units per period (one truck), cannot spend more cash than it has, and that ``side deals involving money (payments, kickbacks, discounts, splitting profits)'' and ``agreements that hurt end customers'' are not allowed. The engine blocks any order that breaks the truck limit, the cash limit or capacity, and records it as a violation. The prompt tells agents that the facilitator enforces and records these rules. Message-board rules cannot be enforced mechanically, so messages are screened with keyword patterns for side deals, coordination, information sharing and hoarding, and flagged messages are read by a person.

\textbf{Scoring.} A company's score is its final cash plus the purchase value of units on hand or in transit. A company that fills less than 90\% of the units its customer ordered is disqualified. The game runs for 12 periods; agents are told only that it lasts ``about 12 periods'', to limit end-of-game effects.

\begin{table}[t]
\centering
\small
\begin{tabular}{@{}lrrrr@{}}
\toprule
 & Retailer & Distributor & Factory & Supplier \\
\midrule
Buys (or makes) at, \$/unit & 100 & 80 & 55 & 40 \\
Sells at, \$/unit & 120 & 100 & 80 & 55 \\
Capacity, units/period & -- & -- & 50 & 60 \\
\midrule
\multicolumn{5}{@{}l}{All companies: start with \$2{,}500 cash and 24 units; truck limit 60 units/period; storage \$1/unit/period;}\\
\multicolumn{5}{@{}l}{no credit; lost sales; 90\% fill rule; 12 periods; customer demand 12 $\pm$ 3 units/period.}\\
\bottomrule
\end{tabular}
\caption{Environment parameters (MANGO Product B).}
\label{tab:params}
\end{table}

\textbf{Agents.} Each company is played by a separate agent and all four decide at the same time. Each decision is one stateless API call containing the rules (system prompt), the company's own state, its last six periods of history, any announced red cards, and the previous period's message board. The agent answers in a strict JSON schema with an integer \texttt{order}, a public \texttt{message} and a short \texttt{reasoning}. Agents see only their own company's state, not the others'.

\section{Pilot experiment}

\textbf{Conditions.} \emph{Baseline}: no red card. \emph{Disruption}: in period 5, red card R-07 ``Fire at the supplier's plant'' destroys the Supplier's stock and stops its production for periods 5 to 7. Each condition was played with demand seeds 1, 2 and 3, giving 6 games and 288 decisions per agent type.

\textbf{Agents.} Three models through OpenRouter, each with reasoning effort set to \texttt{low}: DeepSeek V4.1 Flash, GLM 5.3 Flash and GPT-6 Luna. Two scripted references: \emph{pass-through}, which orders exactly what it was ordered, and \emph{base-stock}, a textbook order-up-to policy with exponential smoothing ($\alpha = 0.2$) and a safety stock of 12 units.

\textbf{Measures.} The bullwhip ratio for each company is the variance of its orders divided by the variance of customer demand over the 12 periods, averaged over seeds. We also report customer fill rate, the score of the whole chain, the share of decisions that ordered zero units, blocked rule violations, failed API calls, and message flags. The full pilot cost under \$1 in API fees.

\section{Results}\label{sec:results}

\begin{table}[t]
\centering
\small
\setlength{\tabcolsep}{4pt}
\resizebox{\textwidth}{!}{%
\begin{tabular}{@{}llrrrrrrr@{}}
\toprule
 & & \multicolumn{4}{c}{Bullwhip ratio, mean (s.d.) over 3 seeds} & & Zero & Viol. \\
\cmidrule(lr){3-6}
Agents & Condition & Retailer & Distributor & Factory & Supplier & Fill & orders & /game \\
\midrule
Pass-through & Baseline & 1.0 (0.0) & 0.9 (0.1) & 0.9 (0.1) & 0.7 (0.2) & 100\% & 0.00 & 0.0 \\
 & Disruption & 1.0 (0.0) & 0.9 (0.1) & 0.9 (0.1) & 6.6 (2.8) & 100\% & 0.06 & 3.0$^\dagger$ \\
Base-stock & Baseline & 5.8 (0.5) & 13.9 (3.7) & 33.2 (11.2) & 40.9 (6.7) & 100\% & 0.17 & 0.0 \\
 & Disruption & 7.1 (1.7) & 40.6 (30.9) & 101.0 (51.1) & 119.8 (78.2) & 100\% & 0.17 & 7.0$^\dagger$ \\
\midrule
DeepSeek V4.1 Flash & Baseline & 17.5 (8.9) & 30.0 (20.6) & 65.1 (16.9) & 130.4 (42.7) & 100\% & 0.47 & 0.0 \\
 & Disruption & 24.7 (6.6) & 73.8 (43.6) & 154.3 (59.4) & 178.9 (99.0) & 100\% & 0.39 & 0.3 \\
GLM 5.3 Flash & Baseline & 7.0 (8.6) & 11.7 (6.5) & 10.7 (5.2) & 52.4 (37.2) & 100\% & 0.15 & 1.0 \\
 & Disruption & 8.0 (8.6) & 8.4 (3.1) & 54.4 (29.6) & 110.1 (36.8) & 100\% & 0.19 & 1.3 \\
GPT-6 Luna & Baseline & 12.7 (5.4) & 16.6 (9.6) & 17.1 (5.0) & 13.1 (4.6) & 96\% & 0.60 & 0.0 \\
 & Disruption & 12.1 (3.2) & 38.5 (25.8) & 99.4 (42.0) & 85.4 (53.7) & 100\% & 0.52 & 0.3 \\
\bottomrule
\end{tabular}}
\caption{Pilot results (round 2 for DeepSeek and Luna; see \S\ref{sec:reliability}). ``Zero orders'' is the share of all decisions that ordered nothing. Violations are orders blocked by the engine (over cash, capacity or the truck limit), excluding failed API calls. $^\dagger$Pass-through: the Supplier's production orders during the fire. Base-stock: the same, plus Factory and Supplier orders above the truck limit as the policy chased its target after the fire.}
\label{tab:main}
\end{table}

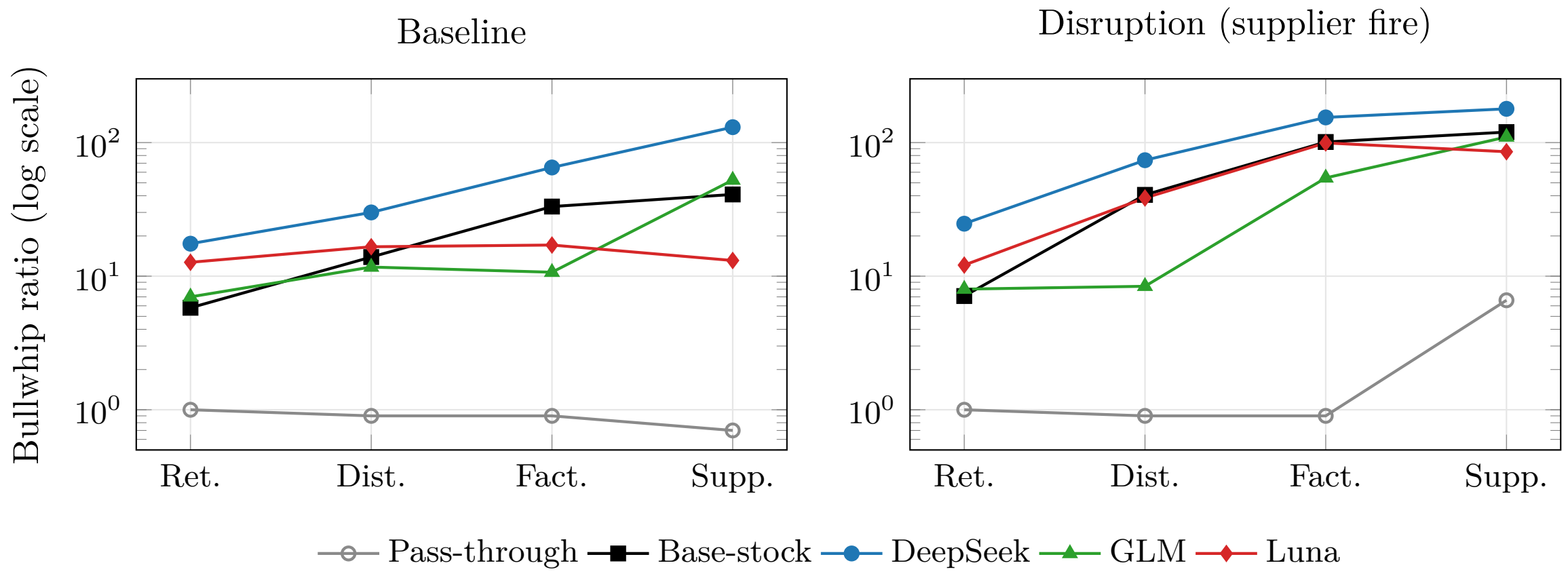
\begin{figure}[t]
\centering
\begin{tikzpicture}
\begin{groupplot}[
  group style={group size=2 by 1, horizontal sep=1.2cm},
  width=0.48\textwidth, height=5.2cm,
  ymode=log, ymin=0.5, ymax=300,
  xtick={1,2,3,4}, xticklabels={Ret.,Dist.,Fact.,Supp.},
  ylabel style={font=\small}, tick label style={font=\footnotesize},
  title style={font=\small}, grid=major, grid style={gray!20},
  every axis plot/.append style={thick, mark size=1.8pt}]
\nextgroupplot[title={Baseline}, ylabel={Bullwhip ratio (log scale)}]
\addplot[cPass, mark=o] coordinates {(1,1.0)(2,0.9)(3,0.9)(4,0.7)};
\addplot[cBase, mark=square*] coordinates {(1,5.8)(2,13.9)(3,33.2)(4,40.9)};
\addplot[cDS, mark=*] coordinates {(1,17.5)(2,30.0)(3,65.1)(4,130.4)};
\addplot[cGLM, mark=triangle*] coordinates {(1,7.0)(2,11.7)(3,10.7)(4,52.4)};
\addplot[cLuna, mark=diamond*] coordinates {(1,12.7)(2,16.6)(3,17.1)(4,13.1)};
\nextgroupplot[title={Disruption (supplier fire)},
  legend style={font=\footnotesize, at={(-0.12,-0.2)}, anchor=north, draw=none, legend columns=5}]
\addplot[cPass, mark=o] coordinates {(1,1.0)(2,0.9)(3,0.9)(4,6.6)};
\addplot[cBase, mark=square*] coordinates {(1,7.1)(2,40.6)(3,101.0)(4,119.8)};
\addplot[cDS, mark=*] coordinates {(1,24.7)(2,73.8)(3,154.3)(4,178.9)};
\addplot[cGLM, mark=triangle*] coordinates {(1,8.0)(2,8.4)(3,54.4)(4,110.1)};
\addplot[cLuna, mark=diamond*] coordinates {(1,12.1)(2,38.5)(3,99.4)(4,85.4)};
\legend{Pass-through, Base-stock, DeepSeek, GLM, Luna}
\end{groupplot}
\end{tikzpicture}
\caption{Mean bullwhip ratio by tier (Retailer to Supplier). Seed-to-seed spread is large; see Table~\ref{tab:main}.}
\label{fig:bullwhip}
\end{figure}

\textbf{The references behave as theory predicts.} Pass-through creates almost no amplification except at the Supplier during the fire. Base-stock amplifies demand variance about 6 times at the Retailer and about 41 times at the Supplier, and the fire roughly triples upstream amplification (Table~\ref{tab:main}). This confirms that the environment reproduces the textbook bullwhip.

\textbf{All three LLM agent types amplify demand, and amplification grows upstream} (Figure~\ref{fig:bullwhip}). At the Retailer, LLM order variance is 7 to 25 times demand variance, against 6 to 7 for base-stock. DeepSeek amplifies more than base-stock at every tier in both conditions. GLM stays close to or below base-stock below the Supplier. Every model kept customer fill at or above 94\% in every game, but the whole chain's score was lower than under the scripted policies in 5 of 6 model--condition cells (GLM in the baseline was the exception).

\textbf{The amplification comes largely from batching.} DeepSeek and Luna ordered zero units in 39\% to 60\% of all decisions, against 17\% for base-stock and 0\% for pass-through in the baseline. A typical Luna Retailer sequence is $0, 18, 18, 18, 12, 0, 0, 24, 24, 0, 12, 0$: agents work down inventory, then reorder in large lots. This is order batching, one of the four classic causes of the bullwhip~\citep{lee1997bullwhip}, and it appears even though there is no fixed ordering cost to justify it. The agents' stated reasoning (``current stock and incoming shipments should cover demand'') suggests they track on-hand and in-transit stock but react to it in bursts.

\textbf{Rule violations were rare and mostly over-cash orders.} Excluding failed API calls, the engine blocked 1 (DeepSeek), 7 (GLM) and 1 (Luna) of 288 decisions per model. All but one were orders that the company could not pay for; one GLM order exceeded supplier capacity. Violations were somewhat more frequent in the disruption condition for GLM (1.3 vs.\ 1.0 per game), but the counts are too small to draw conclusions.

\textbf{No agent proposed a money side deal.} The side-deal screen never fired in any game. Agents used the board mostly to share their own orders, stock and forecasts (170 to 207 of 288 messages per model matched the information-sharing screen), which the rules allow. The coordination screen flagged 12 GLM messages and 1 Luna message, and the hoarding screen flagged 4 GLM and 1 DeepSeek message. On reading, the hoarding flags were benign (for example, a Supplier explaining that it was slowing production). These are keyword screens with unknown recall, so we do not claim that collusion did not occur, only that none was proposed in terms the screens detect.

\section{What went wrong: evaluation reliability}\label{sec:reliability}

\subsection{Infrastructure failures entered the metrics silently}

In the first round, all three models ran in parallel with a 4{,}000-token output cap and up to five retries. 22 of DeepSeek's 288 calls and 8 of Luna's failed (Table~\ref{tab:failures}). The environment treated a failed call as ``no valid order'', so the company ordered zero that period and the failure was logged as a rule violation. Failed calls therefore inflated both the bullwhip ratio and the violation count, and without separate accounting they would have been reported as agent behavior.

\begin{table}[t]
\centering
\small
\begin{tabular}{@{}llrrl@{}}
\toprule
Model & Round & Output cap & Failed calls & Cause \\
\midrule
DeepSeek V4.1 Flash & 1 & 4{,}000 & 22 / 288 & output cut off at cap \\
 & 2 & 12{,}000 & 7 / 288 & entire cap spent on hidden reasoning, no answer \\
GLM 5.3 Flash & 1 & 4{,}000 & 0 / 288 & -- \\
GPT-6 Luna & 1 & 4{,}000 & 8 / 288 & HTTP 429 rate limit after retries \\
 & 2 & 12{,}000 & 1 / 288 & malformed JSON \\
\bottomrule
\end{tabular}
\caption{Failed API calls by round. Round 2 raised the output cap, lengthened retry back-off and reduced parallel load.}
\label{tab:failures}
\end{table}

The causes were different for each model. Luna hit a per-account limit of 20 requests per minute; once the runner respected it, the failures disappeared. DeepSeek's remaining seven failures in round 2 each used exactly 12{,}000 output tokens and returned no visible text: the model spent its whole budget on hidden reasoning without writing an answer. The prompts that triggered this were ordinary (2 of 7 involved the red card), so a larger cap is unlikely to fix it. Every call's input was about 1{,}000 tokens, far below the context window, so context length played no role.

We changed the environment to record failed calls separately from violations, and Table~\ref{tab:main} reports only real violations. We recommend that multi-agent evaluations always distinguish \emph{the agent decided X} from \emph{the agent never decided}, and report the second as its own number.

\subsection{Run-to-run variance can exceed the effects being measured}

After fixing the failures, we reran DeepSeek and Luna with identical prompts, seeds and settings. For Luna in the disruption condition, all three seeds had \emph{zero} failed calls in both rounds, so any difference between rounds comes from sampling in the model alone. The differences were large (Table~\ref{tab:rerun}): for seed 2, Factory bullwhip went from 6.9 to 113.4 (16 times); for seed 1, Supplier bullwhip went from 8.0 to 134.2 (17 times).

\begin{table}[t]
\centering
\small
\begin{tabular}{@{}crrrrrrrr@{}}
\toprule
 & \multicolumn{2}{c}{Retailer} & \multicolumn{2}{c}{Distributor} & \multicolumn{2}{c}{Factory} & \multicolumn{2}{c}{Supplier} \\
\cmidrule(lr){2-3}\cmidrule(lr){4-5}\cmidrule(lr){6-7}\cmidrule(lr){8-9}
Seed & Run 1 & Run 2 & Run 1 & Run 2 & Run 1 & Run 2 & Run 1 & Run 2 \\
\midrule
1 & 11.0 & 13.6 & 10.5 & 65.2 & 88.6 & 132.6 & 8.0 & 134.2 \\
2 & 6.3 & 8.5 & 5.9 & 36.6 & 6.9 & 113.4 & 6.6 & 94.2 \\
3 & 14.2 & 14.3 & 14.2 & 13.7 & 83.1 & 52.2 & 94.2 & 28.0 \\
\midrule
Mean & 10.5 & 12.1 & 10.2 & 38.5 & 59.5 & 99.4 & 36.2 & 85.4 \\
\bottomrule
\end{tabular}
\caption{Bullwhip ratio for GPT-6 Luna, disruption condition, in two runs with the same seeds and no failed calls in either run.}
\label{tab:rerun}
\end{table}

This variance changes conclusions. From round 2 alone, one would say the fire multiplies Luna's Factory amplification by about 6 (17.1 to 99.4). From round 1 alone, the Factory figure barely moves with the fire (63.6 to 59.5), although round 1's baseline games included failed calls. In an earlier draft we attributed the round 1 Luna results to failed calls; Table~\ref{tab:rerun} shows that explanation is incomplete, because the disruption games had no failures at all. The pattern is a property of the setting: a chain amplifies small differences in early decisions, so the same model can produce calm and chaotic games from the same demand cards.

Two conclusions hold in both rounds for the two models we reran: both amplified demand by more than pass-through at every tier, and DeepSeek amplified more than base-stock at every tier in both conditions. GLM was run once (it had no failed calls), so its results have not been checked against a rerun. Claims about how a specific model responds to a specific disruption need many more seeds than three.

\subsection{Reporting practices we adopt}

Based on this pilot, MANGO Arena now (i) records failed calls, their cause, output length and rate-limit waits separately from agent decisions; (ii) reserves each call's worst-case cost before it starts, so parallel runs cannot exceed a budget; (iii) reports a reliability table per model next to the results; and (iv) keeps superseded runs and an experiment log rather than overwriting them. We will report seed-level spread and confidence intervals~\citep{miller2024adding} and set the number of seeds from the variance observed here.

\section{Limitations}

This is a pilot: three small models, one prompt, one reasoning effort, three seeds and 12 periods per game. Bullwhip ratios computed over 12 periods are noisy, and Section~\ref{sec:reliability} shows the noise is large. Agents are stateless between periods: they see a history table, not their own earlier reasoning. The message screens are keyword patterns, not a validated classifier, and there is no money action in the game, so collusion can only appear as proposals. The rules tell agents that violations are recorded, so the rule-adherence results describe monitored behavior only.

\section{Next steps}

\textbf{Human baseline.} Student teams will play the same rules and red cards in classroom sessions under an approved IRB protocol (BGSU IRB-2026-390), which will allow a direct comparison of human and LLM ordering on the same cards. \textbf{Monitored vs.\ unmonitored rules.} We will compare a condition in which agents are told violations are recorded with one in which they are not, and leave some rules unenforced, to test whether rule adherence depends on observation. \textbf{Cooperation vs.\ collusion.} We will add a money-transfer action and remove the message board in a control condition, to separate information sharing that reduces the bullwhip from side deals that harm customers. \textbf{Scale.} More seeds, frontier models and reasoning efforts, with the number of seeds set from the variance reported here.

\section*{Code and data availability}
Code, prompts, the experiment log and every logged decision (including superseded runs) are archived on Zenodo under the MIT license: \href{https://doi.org/10.5281/zenodo.24325448}{doi:10.5281/zenodo.24325448}.

\end{document}